\documentclass[11pt]{article}
\usepackage{natbib,color,fullpage,amsmath,amssymb,amsthm,graphicx,hyperref,listings}

\definecolor{gray}{gray}{.9}

\newcommand{\m}[1]{\mbox{\bf{#1}} }
\newcommand{\etr}{ \ensuremath{ {\rm etr}  }   }
\newcommand{\tr}{ \ensuremath{ {\rm tr}  }   }

\renewcommand{\v}[1]{\mbox{\boldmath{${\rm #1}$}}}

\newcommand{\Exp}[1]{{\rm E}[ \ensuremath{ #1 } ]  }

\newcommand{\Cov}[1]{{\rm Cov}[ \ensuremath{ #1 } ]  }
\lstnewenvironment{rcode}
 {\lstset{basicstyle=\small,stringstyle=\ttfamily,backgroundcolor=\color{gray}}}
 {}

\begin{document}

\title{Hierarchical multilinear models for multiway data}
\author{Peter D. Hoff \thanks{Departments of Statistics and Biostatistics , 
University of Washington,
Seattle, Washington 98195-4322.
Web: \href{http://www.stat.washington.edu/hoff/}{\tt http://www.stat.washington.edu/hoff/}. The author thanks an associate editor and two reviewers whose comments improved the content and style of this article. 
This work was partially supported by NSF grant  SES-0631531. 
        }}
\date{ \today }
\maketitle

\begin{abstract}
Reduced-rank decompositions provide descriptions  of  the
variation among the elements of a matrix or array. 
In such decompositions, the elements of an array are expressed 
as products of low-dimensional latent factors.
This article
presents a model-based version of such a decomposition, 
extending the scope of reduced rank methods to accommodate  a variety of 
data types such as 
longitudinal social networks  and 
continuous  multivariate data that are  cross-classified by 
categorical variables. 
The proposed model-based approach is hierarchical, in that the 
latent factors corresponding to a given dimension of the array 
are not {\it a priori} independent, but exchangeable. Such 
a hierarchical approach allows more flexibility in the types of patterns 
that can be represented. 

\medskip 

\noindent {\it Some key words}:  Bayesian, multiplicative model, PARAFAC, regularization, shrinkage. 
\end{abstract}

\section{Introduction}
Matrix-valued data are prevalent in many scientific disciplines. 
Studies in  social and health 
sciences  often gather social network data 
that can be represented by 
square, binary matrices with  undefined diagonals. 
Numerical results from gene expression studies are
recorded in matrices with 
rows  representing tissue samples and columns representing genes. 
Analysis of stock market returns involves data matrices with rows 
representing stocks and  columns representing time. 
With such data
there are often dependencies among both the rows and the columns 
of the data matrices, and so 
 the standard tools of multivariate analysis, in which patterns along one 
dimension of the data matrix are thought of as i.i.d., may be inadequate
for data analysis purposes. 
As an alternative to the i.i.d.\ paradigm, patterns of row and column variation in matrix-valued data are often 
described with reduced-rank matrix decompositions and models. 
For example, the $i,j$th entry of an $m_1\times m_2$ 
matrix might be expressed as 
$y_{i,j} = \langle \v u_i, \v v_j \rangle + \epsilon_{i,j}$, where 
the heterogeneity among a set of low-dimensional vectors 
$\{\v u_1,\ldots,\v u_{m_1}\}$ and $\{\v v_1,\ldots,\v v_{m_2}\}$ is used to represent heterogeneity attributable  to the 
row and column objects respectively. 
Such models can be described as being
bilinear, as the expectation of $y_{i,j}$ 
is a bilinear function of the parameters. These models are related to 
biplots \citep{gabriel_1971}, bilinear regression \citep{gabriel_1998}
and the singular value decomposition (SVD). 

In more complex situations  the data
take the form of a multidimensional array 
instead of a matrix. For example, temporal variation in a
social network over a discrete set of time points may be represented
by a three-way array $\v Y=\{ y_{i,j,t} \}$, where $y_{i,j,t}$
describes the relationship between nodes $i$ and $j$ at time $t$.
Similarly, gene expression data gathered under a variety of experimental 
conditions, or multiple variables measured on a set of companies over time
are also examples of array-valued or multiway data. 
Surveys of multiway data analysis include \citet{coppi_bolasco_1989}
and 
\citet{kroonenberg_2008}. 
The July-August  2009 issue of the Journal of 
Chemometrics was dedicated to Richard Harshman, one of the founders of 
three-way data analysis. 
Harshman \citep{harshman_1970,harshman_lundy_1984} 
developed a three-way generalization of the  
SVD known as ``parallel factor analysis'', or PARAFAC, that 
has become one of the primary methods of multiway data analysis. 
%The generalization is as follows:
While the SVD represents the
$i,j$th element of a rank-$R$ matrix $\m A$ as $a_{i,j} = \langle \m u_i,\m v_j \rangle \equiv\sum_{r=1}^R u_{i,r} v_{j,r}$, 
the PARAFAC decomposition of a three-way array
represents the $i,j,k$th element as $a_{i,j,k} = \langle \m u_i,\m v_j , \m  w_k\rangle  =\sum_{r=1}^R 
 u_{i,r} v_{j,r} w_{k,r}$. 
\citet{kruskal_1976,kruskal_1977} related such decompositions to a 
precise definition of rank for three-way arrays,
in which the rank is the smallest integer $R$ for which the above 
representation holds. The generalization to arbitrary dimensions is 
straightforward:
A $K$-dimensional array of rank $R$ is one in which the elements can  
be expressed as 
a multilinear function of  $R$-dimensional factors. 
A compact review of 
these results and others appears in \citet{kruskal_1989}. 

While the area of multiway data analysis has been active, most of the 
focus has been on 
algorithms for
finding least-squares solutions, pre- and post-processing of results, and 
interpretation of the least-squares parameters. 
Little has been done in terms 
of incorporating  multilinear representations into 
statistical  models. 
One exception is the work of
\citet{vega_2003} and \citet{vega_2005}, who develop
algorithms for finding maximum likelihood solutions for situations 
with heteroscedastic  or correlated error terms. However, these algorithms 
assume the error variance is known. 

This article develops a hierarchical multilinear model for incorporation into
a variety of non-standard multiway data analysis situations, and presents
a Bayesian approach for parameter estimation. 
The motivation is twofold: First, 
multilinear array representations can involve a large number of parameters. 
Overfitting of the model can be ameliorated by using shrinkage 
estimators provided by a Bayesian approach. In particular, 
a hierarchical Bayesian approach can be used 
to provide shrinkage patterns that are based primarily 
on the observed data, rather 
than relying heavily on a fixed prior distribution. 
The second motivation is that Bayesian approaches and MCMC estimation 
methods allow one to incorporate the basic multilinear representation  into 
models for complex data that might involve 
additional dependence structures or discrete data. 

After presenting the hierarchical multilinear model and Bayesian methods 
for estimation in Sections 2 and 3, 
a small simulation study is presented in Section 4 
to compare mean squared errors
of three different parameter estimation methods:
least-squares, a simple
non-hierarchical Bayesian approach and a Bayesian hierarchical approach.
The Bayes estimators are found to outperform the least-squares estimator,
with the hierarchical Bayes procedure giving the best performance.
The performance of the estimators when the rank
of the model is misspecified is also considered. In this situation, the
least-squares and non-hierarchical Bayes procedures increasingly overfit
the data as the rank is increased, while the hierarchical Bayes procedure
is robust to rank misspecification.

Sections 5 and 6 give examples in which it is useful to embed a multilinear 
model within a larger model for observed data. 
Section 5 considers estimation of a multivariate mean $\Exp{ \v y_{\rm x}}= \v \mu_{\rm x} $ for each possible value of a vector of categorical 
variables  $\v x$. 
Often the number of observations per level of $\v x$ is 
small and varies from level to level. 
A hierarchical model for the mean,  
$\v \mu_{\rm x}  \sim$  multivariate normal$( \v \beta_{\rm x}, \v\Sigma)$, 
allows for consistent estimation of each $\v \mu_{\rm x}$ but shrinkage towards
$\v \beta_{\rm x}$ when the sample size is small. The values 
$\m B =\{\v \beta_{\rm x} : \v x\in \mathcal X\}$ can be represented as a 
multiway array,  and a reduced rank multilinear model for $\m B$ allows 
for the modeling of non-additive effects of $\v x$ with a relatively small 
number 
of parameters. 

Section 6 presents an analysis of international cooperation 
and conflict  during the cold war.  The data consist of a three-way 
array with element $y_{i,j,t}$ representing the relationship between 
countries $i$ and $j$ in year $t$. 
Several features of these data make existing tools from multiway data 
analysis  inappropriate, one being that 
the data are ordinal. The range of the data 
includes the integers from -5 to 2, indicating different levels 
of military cooperation or conflict. Assuming that the $y_{i,j,k}$'s are 
normally 
distributed or even continuous would be inappropriate. 
However,  using the tools developed in this article 
it is reasonably straightforward to embed a 
multilinear representation within an ordered probit model for these data. 
A discussion of the results and directions for future research 
follows in Section 7.

\section{Reduced rank models for array data}
In this section we review the reduced rank model and an alternating least-squares (ALS) procedure for parameter estimation. For a review of the properties, limitations  and alternatives to ALS,
see  \citet{tomasi_bro_2006} and 
Chapter 5 of \citet{kroonenberg_2008}.

\subsection{Rank and factor representations for arrays}
Given an $m_1\times m_2$ data matrix $\m Y$ it is often desirable to separate out the 
``main features'' of $\m Y$ from the ``patternless noise.''
This motivates a model of the form $\m Y = \v \Theta + \m E$, where 
$\v \Theta$ is to be estimated from the data. Interpreting ``main features''
as those that can be well-approximated by a low-rank matrix, the 
rank of $\v \Theta$ is usually taken to be some value $R< m_1\wedge m_2$. 
The rank of a matrix $\v \Theta$
can be defined as the smallest integer $R$ such that
there exists matrices $\m U\in \mathbb R^{m_1\times R}$ and $\m V\in \mathbb R^{m_2\times R} $ such that 
%\begin{eqnarray*}
%\v \Theta &=& \sum_{r=1}^R \v u_r \otimes \v v_r = \m U \m V^T 
%    \ , \ \mbox{or equivalently,} \
%\theta_{i,j}=\sum_{r=1}^R u_{i,r} v_{j,r}  =  \langle \v u_i ,\v v_j \rangle, 
%\end{eqnarray*}
%where
%$\v u_r$ is the $r$th column of $\m U$ in the first equation and 
% $\v u_i$  is the $i$th row of $\m U$ in the second. 
\[ 
\v \Theta = \sum_{r=1}^R \v u_r \otimes \v v_r = \m U \m V^T , 
\]
where $\v u_r$ is the $r$th column of $\m U$, or equivalently 
\[
\theta_{i,j}=\sum_{r=1}^R u_{i,r} v_{j,r}  =  \langle \v u_i ,\v v_j \rangle, 
\]
where $\v u_i$  is the $i$th row of $\m U$. 
Variation among the rows of $\m U$
 represents  the heterogeneity 
in $\v \Theta$ attributable to variation in the row objects, 
and similarly
variation among the rows of 
$\m V$ represents heterogeneity attributable to 
the column objects. 

A $K$-order multiway array $\m Y$  with dimension 
$m_1\times \cdots \times m_K$
has elements 
$\{ y_{i_1,\ldots, i_K} :  i_k \in \{ 1,\ldots,m_k\}  \}$. 
As with a matrix, we may define a model for a $K$-order array 
as $\m Y=\v \Theta + \m E$, where $\m E$ is an array of uncorrelated, 
mean-zero noise and $\v \Theta$ is a reduced rank array to be estimated. 
Following \citet{kruskal_1976} and \citet{kruskal_1977}, 
the rank of a $K$-order array $\v \Theta$ is simply 
the smallest integer $R$ such that there exist 
matrices $\{ \m U^{(k)} \in \mathbb R^{m_k\times R}, k=1,\ldots, K\}$, such 
that
%\begin{eqnarray*}
%\v \Theta  &=& \sum_{r=1}^R \v u^{(1)}_r \otimes \cdots \otimes \v u^{(K)}_r 
%   \equiv \langle  \m U^{(1)} ,\ldots, \m U^{(K)} \rangle \ , \ \mbox{ or equivalently} \\
%\theta_{i_1,\ldots, i_K} &=& \sum_{r=1}^R u^{(1)}_{i_1,r} \times 
%  \cdots \times  u^{(K)}_{i_K,r}  
% \equiv 
%  \langle \v u_{i_1}^{(1)} , \cdots,  \v u_{i_K}^{(K)} \rangle,
%\end{eqnarray*}
%where $\v u_r^{(k)} $ is the $r$th column 
%of $\m U^{(k)}$ in the first equation and 
% $\v u_{i}^{(k)} $ is the $i$th row of $\m U^{(k)}$ in the second. 
\[ 
\v \Theta  = \sum_{r=1}^R \v u^{(1)}_r \otimes \cdots \otimes \v u^{(K)}_r 
\equiv \langle  \m U^{(1)} ,\ldots, \m U^{(K)} \rangle ,
\]
where $\v u_r^{(k)} $ is the $r$th column of $\m U^{(k)}$, or equivalently
\[ 
\theta_{i_1,\ldots, i_K}=\sum_{r=1}^R u^{(1)}_{i_1,r} \times \cdots \times  u^{(K)}_{i_K,r} \equiv \langle \v u_{i_1}^{(1)} , \cdots, \v u_{i_K}^{(K)}\rangle,
\]
where $\v u_{i}^{(k)} $ is the $i$th row of $\m U^{(k)}$. 
As in the matrix case, variation among the rows of 
$\m U^{(k)}$ represents heterogeneity attributable to the $k$th 
set of objects, that is, the $k$th mode of the array. 

\subsection{Least squares estimation}
In the matrix case the  least squares estimate of $\v \Theta=\m U\m V^T$ 
(also the MLE assuming normal, i.i.d.\ 
errors) can be obtained from the 
first $R$ components of the singular value decomposition of $\m Y$. 
For arrays of higher order, only iterative methods of estimation are
available. 
Perhaps the simplest method of parameter estimation is the
alternating least squares algorithm (ALS), in which factors corresponding
to a given mode are updated to minimize the residual sums of
squares given the current values for the other modes.
In this subsection we review the relevant calculations for ALS, which will 
also be useful for Bayesian estimation  in the next section. 

\paragraph{Estimation for a three-way model:}
We begin with a three-way array
so that the main ideas can be understood with a minimal amount of 
notational complexity. 
Let $\m Y$ be a three-way array 
modeled as  $y_{i,j,k}
= \langle\v u_{i} ,\v v_{j} , \v w_{k} \rangle  + \epsilon_{i,j,k}$,  with $\{ \epsilon_{i,j,k} \} \sim$ i.i.d.\ 
normal$(0,\sigma^2)$. 
We can write 
\begin{eqnarray*} 
\v y_{i,j,\cdot}  &=& \v W (\v u_i \circ \v v_j ) + \v \epsilon_{i,j,\cdot} \\
\v y_{i,\cdot,k}  &=& \v V (\v u_i \circ \v w_k ) + \v \epsilon_{i,\cdot,k} \\
\v y_{\cdot,j,k}  &=& \v U (\v v_j \circ \v w_k ) + \v \epsilon_{\cdot,j,k}, 
\end{eqnarray*}
where $\m U$, $\m V$, $\m W$ are $m_1\times R$, $m_2\times R$ and $m_3\times R$ matrices respectively, 
$\v u_i$, $\v v_j$, $\v w_k$ are rows of these matrices, 
$\v y_{i,j,\cdot}$, $\v y_{i,\cdot,k}$, 
 $\v y_{\cdot,j,k}$ are vectors of length $m_1$, $m_2$ and $m_3$, and 
``$\circ$'' denotes the Hadamard product (elementwise multiplication).
Some matrix algebra and careful summation shows that, as a function of 
$\m U$, $p(\m Y| \m U, \m V, \m W)$ can be written 
\begin{eqnarray}
p(\m Y| \m U, \m V, \m W) &\propto& 
\etr (  \v U^T \v L/\sigma^2 - \v U^T \v U \v Q/[2\sigma^2]) \ , \mbox{where}\\ \label{eq:lcu1fc}
\v Q &=& (\m V^T \m V )\circ (\m W^T\m W )  \ \  \mbox{and } \nonumber \\
\v L &=&  \sum_{j,k} \v y_{\cdot,j,k} \otimes ( \v v_j \circ \v w_k )  \ , \nonumber
\end{eqnarray}
and $\etr(\m A) = \exp\{ {\rm trace}(\m A)\}$.
With $\m V$ and $\m W$ fixed, the conditional MLE and least-squares estimate of 
$\m U$ is given by $\hat {\m U} = \m L \m Q^{-1} $. 
The ALS procedure is to iteratively replace a current value of $\v U$ with its 
conditional least-squares estimate, then replace $\v V$ and $\v W$ similarly.
This procedure is then iterated until a pre-specified convergence criterion has been met. Typically, the algorithm is replicated beginning with  several different randomly generated initial values, with each replicate iterated until the relative fit 
 $|| \m Y - \hat {\v \Theta}||^2/||/||\m Y||^2$ is below a user-defined 
threshold.  
A comparative study of different least-squares estimation methods 
done by \citep{tomasi_bro_2006}  concluded that the ALS procedure 
provides a good compromise between computational complexity and 
quality of the solution.

\paragraph{Estimation for a $K$-way model:}
Now suppose $\m Y$ is an $m_1\times \cdots\times m_K$ array. 
Let $\m U^{(1)},\ \ldots,\ \m U^{(K)}$
be the matrices of factors for the $K$ modes, so that $\m U^{(k)}$ is an 
$m_k\times R$ matrix.
%The basic results from the three-way model carry over as follows:
Generalizing the approach for the three-way model, 
let $\v y_{\mathbf i_{1}} = ( y_{1,i_2,\ldots, i_K},\ldots, 
   y_{n_1,i_2,\ldots, i_m} )$ be a ``fiber'' along the first dimension 
of the array. 
Then  
we can write $\m y_{\mathbf i_{1}} = \m U^{(1)}  ( \m u^{(2)}_{i_2} \circ{\m u^{(3)}_{i_3}} \circ \cdots \circ {\m u^{(K)}_{i_K}} ) + \epsilon_{\mathbf i_{1}}.
$
Similar to the three-mode case, 
as a function of
$\m U^{(1)}$, $p(\m Y| \m U^{(1)}, \ldots, \m U^{(K)} )$ can be written
\begin{eqnarray}
p(\m Y| {\m U^{(1)}},  \ldots, {\m U^{(K)}}) &\propto&
 \etr ( {\m U^{(1)}}^T \v L/\sigma^2 - {\m U^{(1)}}^T {\m U^{(1)}} \v Q/[2\sigma^2]) \ ,
 \mbox{where} \label{eq:lcufc} \\
\v Q &=& ({\m U^{(2)}}^T {\m U^{(2)}} )\circ \cdots \circ ({\m U^{(K)}}^T {\m U^{(K)}} )  \ \mbox{and } \nonumber \\
\v L &=&  \sum_{i_2,\ldots,i_m} \v y_{\mathbf i_{1}} 
  \otimes ( {\m u^{(2)}}\circ \cdots \circ {\m u^{(k)}}  )  \ . \nonumber
\end{eqnarray}
The conditional MLE and least squares estimator of $\m U^{(1)}$ given 
the factor values for the other modes is thus $\hat {\m U}^{(1)} = 
 \m L\m Q^{-1}$. As with three-way data, the ALS procedure is to 
iteratively replace the factors matrices with  their 
conditional least-squares estimates until convergence. 

\section{Bayes and hierarchical Bayes estimation}
Compared to least-squares or maximum likelihood methods,
Bayesian procedures often provide stable estimation in high-dimensional
problems due to regularization via the prior 
distribution. Using conjugate prior distributions, 
this section provides a 
Gibbs sampling scheme 
that approximates 
the posterior  distribution $p( \v U^{(1)},\ldots, \v U^{(K)},\sigma^2 | \m 
Y  )$, and by extension, an approximation to the         
posterior distribution of $\v \Theta = \langle \v U^{(1)},\ldots, \v U^{(K)} \rangle $. 
The posterior expectation of $\v \Theta$ can be used as a
Bayesian estimate of the main features of the data array. 
\subsection{A basic Gibbs sampler}
Let the prior distribution for $\m U^{(k)}$ be such that 
 that the rows of $\m U^{(k)}$ are i.i.d.\ multivariate normal$(\v \mu_k,\v\Psi_k)$ or equivalently, 
$\m U^{(k)} \sim $ matrix normal$(\m M_k =\v 1 \v \mu_k^T  ,\v\Psi_k,\m I)$ with density
\begin{eqnarray*}
p(\v U^{(k)}) & \propto & \etr( -(\m U^{(k)} - \m M_k)^T (\m U^{(k)} -\m M_k) \v \Psi_k^{-1}/2)  \\
 &\propto&   \etr( \m U^{(k)T} \m M_k  \v \Psi^{-1}_k -  \m U^{(k)T} \m U^{(k)} \v \Psi_k^{-1}/2). 
\end{eqnarray*}
Combining this with the likelihood from Equation \ref{eq:lcufc}, it follows 
that 
if ${\m U^{(1)}}\sim$ matrix normal$(\v M_1, \v \Psi_1,\m I)$ 
{\it a priori}, then
the full conditional distribution is also matrix normal with density
\begin{eqnarray*}
p({\m U^{(1)}} | \v Y, {\m U^{(2)}},\ldots, {\m U^{(K)}} ) &\propto  &
  \etr( -({\m U^{(1)}} - \tilde{\m M}_1)^T ({\m U^{(1)}} -\tilde {\m M}_1) /2)\tilde
 {\v \Psi}_1^{-1})
 \\ \label{eq:mmnfc}
\tilde {\v \Psi}_1 &=& (  \v Q/\sigma^2+ \v \Psi^{-1}_1 )^{-1}    \nonumber \\
\tilde {\v M}_1 &=&  (  \m L/\sigma^2 + \v M_1 \v \Psi^{-1}_1 )\tilde{\v \Psi}_1 \nonumber.
\end{eqnarray*}
Full conditional distributions for $\m U^{(2)},\ldots, \m U^{(m)}$ are derived 
analogously.  
Using a conjugate inverse-gamma$(\nu_0/2,\nu_0\sigma_0^2/2)$ prior
distribution for $\sigma^2$ results in an inverse-gamma$(a,b)$  full conditional
 distribution where $a=(\nu_0+\prod_k m_k)/2$ and
$b= (\nu_0\sigma_0^2 + ||\m Y - \langle \m U^{(1)} ,\ldots, \m U^{(K)} \rangle
||^2)/2$.

A Markov chain Monte Carlo approximation to
$p( \v U^{(1)},\ldots, \v U^{(K)},\sigma^2 | \m Y )$
can be made by iteratively sampling each unknown quantity
from its full conditional distribution.
This generates a Markov chain, samples from which converge in distribution 
to $p( \v U^{(1)},\ldots, \v U^{(K)},\sigma^2 | \m Y )$. 
However, it would be inappropriate to  estimate 
$\v U^{(k)}$ by its posterior mean $\hat {\v U}^{(k)}$, or 
$\v \Theta$ with 
$\langle \hat {\v U}^{(1)},\ldots, \hat{\v U}^{(K)} \rangle$,  as 
the  values of the latent factors are not separately
identifiable. For example, 
the 
likelihood is invariant to joint permutations and complementary rescalings of the columns of the 
$\m U^{(k)}$'s (see \citet{kruskal_1989} for a discussion of  
the uniqueness of reduced-rank array decompositions). 
 Instead, the posterior 
mean estimate $\hat {\v \Theta}$  of $\v \Theta$, obtained from 
the average of 
 $\langle  {\v U}^{(1)},\ldots, {\v U}^{(K)} \rangle$ over iterations of the 
Markov chain, can be used as a point estimate of $\v \Theta$.  If desired, 
point estimates of the $\m U^{(k)}$'s can then be obtained 
from a rank-$R$ least-squares approximation of $\hat {\v \Theta}$.

\subsection{Hierarchical modeling of factors}
%Although certain aspects of the $\m U^{(k)}$ are not identifiable from the 
%data, other features manifest themselves as patterns in the array $\v \Theta$. 
%For example,  
Rarely will we have detailed prior knowledge of 
an appropriate mean $\v \mu_k$ and variance $\v \Psi_k$ 
for each factor matrix $\m U^{(k)}$. Absent these, we may consider 
a simple ``weak'' prior distribution such as 
$\v u^{(k)}_1,\ldots, \v u^{(k)}_{m_k} \sim $ i.i.d.\ 
multivariate normal$(\v 0, \tau^2 \m I$), where $\tau^2$ is 
large. However, doing so would ignore patterns of heterogeneity in 
the $\v U^{(k)}$'s that could improve estimation of $\v \Theta$.  Even though 
the value of $\v \Theta$ is invariant to certain sign changes or 
permutations of the column of $\{ \m U^{(k)}, k=1,\ldots, K\}$, 
other patterns in the $\m U^{(k)}$'s manifest themselves as patterns
in $\v \Theta$ and $\m Y$. 

To illustrate this, recall that the factors 
represent variance among the elements of the data array $\m Y$  that can 
be attributed to heterogeneity within the various modes. %To illustrate, 
Consider three mode data in which the first mode represents a large number of experimental 
units and the other two modes represent two sets of experimental conditions. 
In this case, $y_{i,j,k}$ is the measurement for unit $i$ when condition 
one is at level $j$ and condition two is at level $k$. 
Letting the factors corresponding to the three modes be $\m U$, $\m V$ and $\m W$, 
modeling 
the rows $\v u_1,\ldots, \v u_{m_1}$
of the  $m_1\times R$ factor matrix $\v U$ 
as i.i.d.\ multivariate normal$(\v \mu,\Psi)$ induces a covariance among 
the elements of each unit-specific  $m_2\times m_3$ matrix 
$\v Y_i = \{ y_{i,j,k}, 1\leq j\leq m_2,1\leq k\leq m_3\}$, given by the 
following calculation:
\begin{eqnarray}
\v u_i &=& \v\mu + \v \gamma_i , \  \v \gamma_i\sim\mbox{multivariate normal}(\v 0,\v \Psi) \nonumber \\
y_{i,j,k} &=& \langle \v u_i ,\v v_j ,\v w_k \rangle + \epsilon_{i,j,k}  \nonumber \\
      &=&  \v u_i^T ( \v v_j \circ \v w_k ) + \epsilon_{i,j,k} 
   = \v \mu^T ( \v v_j \circ \v w_k )  + \v \gamma_i^T ( \v v_j \circ \v w_k ) + \epsilon_{i,j,k}   \label{eq:hmean} \\
\Cov{ y_{i,j,k} ,y_{i,l,m} } &=&  
\Exp{\v\gamma_i^T (\v v_j \circ \v w_k) (\v v_l \circ \v w_m)^T \v \gamma_i} +\sigma^2 \m I \nonumber \\
&=& \tr( (\v v_j \circ \v w_k) (\v v_l \circ \v w_m)^T \v \Psi )  +\sigma^2 \m I \nonumber \\
 &=& \tr( [  (\v v_j \v v_l^T ) \circ (\v w_k \v w_m^T ) ] \v \Psi)  + \sigma^2 \m I. 
  \label{eq:hvar}
\end{eqnarray}
Each unit has a measurement under conditions $(j,k)$  and under $(l,m)$, 
and the correlation of these measurements across experimental units 
is 
determined by  $\v v_j\v v_l^T$ ,
$\v w_k\v w_m^T$ and the covariance matrix $\v \Psi$. 
Additionally, 
Equation \ref{eq:hmean} indicates that  the scale  of $\v \mu$ 
relative to that of $\v \Psi$ represents how much variability there is 
among the  units. 
%elements of $\v \Theta $ along a given dimension. 
Fixing 
$\v \mu$ or $\v \Psi$ in advance 
places restrictions on these variances and correlations. 
This suggests the use of a hierarchical model as an alternative, whereby 
the 
mean and variance of the 
factors of each mode are estimated from the observed data. 
Returning to the general case of $K$ modes, the proposed hierarchical model 
 is 
as follows:
\begin{eqnarray*}
 \{ \v u^{(k)}_1,\ldots, \v u^{(k)}_{m_k} \}   &\stackrel{\rm iid}{\sim} &
  \mbox{multivariate normal}( \v \mu_k ,\v \Psi_k  )  \\
 \v \mu_k | \v \Psi_k &\sim& \mbox{multivariate normal}(\v\mu_0 ,\v 
\Psi_k/\kappa_0) \\
 \v \Psi_k  & \sim& \mbox{inverse-Wishart}( \m S_0, \nu_0   ) . 
\end{eqnarray*}

Readers familiar with factor models for matrices (the case of $K=2$)  may be 
concerned about the non-orthogonality of the columns of the latent factor 
matrices in the above model. In the matrix case, the mean matrix 
for $\m Y$ is given by $\v \Theta =\m U^{(1)} \m U^{(2)T}$. 
Letting $\tilde {\m U}^{(k)}  = \m U^{(k)} \m H$, $k=1,2$
we see that $\v \Theta =\tilde{ \m U}^{(1)}\tilde{ \m U}^{(2)T}$
 for any
orthonormal matrix $\m H$. This invariance to rotation in the matrix 
case, however, does not generalize to 
rotation invariance  for multilinear representations of arrays.
\citet{kruskal_1977} shows that other 
than some elementary invariances (such as 
a common relabeling of the columns of 
all the factor matrices), 
multilinear factor representations are generally rotationally unique.

Diffuse priors  can be used as a default, such as $\v \mu_0 = \m 0$, $\kappa_0=1$, $\nu_0=R+1$ and 
$\m S_0^{-1}= \m I\tau^2_0$, where $\tau^2_0$ is some pre-specified value determined by the scale of the measurements. 
As an alternative, 
unit information prior distributions \citep{kass_wasserman_1995} can be used, which 
weakly center the prior parameters around estimates obtained from the data. 
For example, $\tau_0^2$ could be obtained as the variance of 
latent factor estimates obtained from a rank-$R$ least squares approximation to
$\m Y$, and the prior distribution for $\sigma^2$ could be weakly 
centered around 
the corresponding residual variance. 
In either case, the full conditional distributions for all parameters have straightforward
derivations, and are summarized in the following Gibbs sampling scheme:
Given current values of $\{ \m U^{(1)},\ldots, \m U^{(K)} \}$ and $\sigma^2$, 
new values of these parameters are generated as follows:
\begin{enumerate}
\item For each $k \in \{1,\ldots, K\}$ in random order, 
\begin{enumerate}
\item sample $\v \Psi_k \sim $ inverse-Wishart$( [\m U^{(k)T}\m U^{(k)}+ \m I\tau^2_0]^{-1}, m_k+ R+1 )$; 
\item sample $\v \mu_k \sim $ multivariate normal$( \m U^{(k)T}\v 1/[m_k+1]  ,\v \Psi_k/[m_k+1])$; 
\item sample $\m U^{(k)}\sim $ matrix normal$(\tilde{\m M}_k,
  \tilde{\v \Psi}_k,\v I)$, where
\begin{itemize}
\item $\tilde{\v \Psi}_k = (  \v Q_k/\sigma^2+ \v \Psi_k^{-1} )^{-1}  $, and
\item $\tilde{\v M}_k=(\m L_k/\sigma^2+\v 1 \v \mu_k^T \v \Psi_k^{-1} )\tilde{\v \Psi}_k$. 
\end{itemize}
\end{enumerate}
\item Sample $\sigma^2 \sim $ inverse-gamma$(  \tilde \nu_0/2,\tilde \nu_0\tilde \sigma^2_0/2)$, where
\begin{itemize}
\item $\tilde \nu_0 =  \nu_0 + \prod_{k=1}^K m_k $, and
\item  $\tilde \nu_0 \tilde  \sigma_0^2 =  \nu_0 \sigma^2_0 + 
     || \m Y - \langle \m U^{(1)},\ldots,\m U^{(K)} \rangle ||^2$. 
\end{itemize}
\end{enumerate}
Note that $\{(\v \mu_k, \v \Psi_k),k=1,\ldots, K \}$ will not 
be separately identifiable since, for example, the scales of $\{\v U^{(k)},k=1,\ldots, K\}$ are not separately identifiable.  
However,  a non-hierarchical Bayesian approach 
restricts the overall scale of $\v \Theta$, 
as well the shrinkage point for the $\m U^{(k)}$'s. 
In contrast, the hierarchical model allows 
these things to be determined by the data. 

\section{Comparison of estimators}This section presents the results of some simulation studies comparing the performance of the hierarchical Bayes
 procedure to ALS estimation.
In the first study,
one-hundred random $\v\Theta$-arrays were generated, each having dimension
$m_1\times m_2\times m_3 =10\times 8\times 6$ and  rank $R=4$,
and each to be estimated
from a corresponding ``observed'' data array $\m Y$.
Letting $\tilde R = m_2 \times m_3$,
the $\v\Theta$ and $\m Y$ arrays were generated as follows:
\begin{enumerate}
\item For each mode $k\in \{1,2,3\}$,
\begin{enumerate}
\item  sample $\v\Psi_k$ as follows:
\begin{enumerate}
\item sample $\v\Psi_0 \sim $ Wishart $(\m I, \tilde R+1 )$,
\item set $\nu_0 =\tilde R+x$ where $x\sim$ Poisson($\sqrt{\tilde R}$),
\item sample $\v\Psi_k \sim$ inverse-Wishart$(\v\Psi_0, \nu_0)$;
\end{enumerate}
\item sample $\v \mu_k \sim $ multivariate normal $(\v 0, \v \Psi_k)$; 
\item sample $\tilde {\v U}^{(k)} \sim $ multivariate normal $(\v \mu_k, \v\Psi_k)$.
\end{enumerate}
\item  Let $\v \Theta$ be the rank-$R$ least-squares approximation to      $\langle \tilde {\v U}^{(1)},\tilde{\v U}^{(2)} , \tilde {\v U}^{(3)} \rangle$, but rescaled so that the
average squared magnitude of the elements $\sum \theta_{i,j,k}^2/(m_1 m_2 m_3)$ is 1.
\item Set $\v Y= \v \Theta + \v E$,
   where $\{ \epsilon_{i,j,k} \} \stackrel{\rm iid}{\sim}$ normal$(0,1/4)$.
\end{enumerate}
We now go through the rationale
for this simulation scheme.
Working backwards, in steps 2 and 3 the
error variance for $\m E$ is set to be 1/4 of the average squared magnitude of the elements of $\v \Theta$.  This makes estimation of  $\v \Theta$ feasible but
 not trivial.
In steps 1 and 2,
we first generate an array having a maximal rank $\tilde R$, and then
let $\v \Theta$ be its rank-4 least-squares approximation.
The  rationale for this is to make the generated $\v \Theta$ arrays
somewhat different in distribution from the prior distribution that
is used for estimation. %, thus giving a more fair comparison
%between the performance of the Bayesian procedure and ALS estimation.
If instead the parameter values were simulated from the prior 
distribution used for estimation, we would expect the Bayes procedure to 
outperform the ALS procedure simply because the the Bayes estimates would be
{\it a priori} weighted towards their true values. 
By generating $\v \Theta$ from a distribution other than the prior, 
we intend to give a  more fair comparison
between the performance of the Bayesian procedure and ALS estimation.
Additionally, the ``prior'' parameters $\v \Psi_0$ and $\nu_0$
in steps 1.(a) are randomly generated in order to provide a broader
range of patterns generated in the $\v\Theta$ arrays than could be obtained
from fixed values of $\v \Psi_0$ and $\nu_0$.

\subsection{Known rank}
We first examine the case where the
presumed rank of $\v \Theta$
is equal to the true rank of 4.
Two estimates were computed
for each of the one-hundred simulated $\v \Theta$-arrays:
\begin{itemize}
\item[]$\hat {\v \Theta}_{\rm LS}$ (least squares), an estimate obtained via the alternating least-squares algorithm;
\item[] $\hat {\v \Theta}_{\rm HB}$ (hierarchical Bayes), a posterior  estimate under the hierarchical
model and unit information priors described in Section 3.2. 
\end{itemize}

The least squares estimates were obtained by running the ALS algorithm using
twenty different random starting values
and then selecting the one
that gave the minimum residual sum of squares. For each starting value,
the ALS algorithm was iterated until the magnitude of the change
in the estimate, relative to the magnitude of the estimate, was less than
$10^{-6}$.

The Bayesian estimates were obtained using the Gibbs sampling scheme
described in the previous section, with 1,000 iterations
to allow for convergence to the stationary distribution (``burn-in''),
followed
by 10,000 iterations for estimating the mean matrix.
Mixing of the algorithm was assessed by monitoring the
value of $|| \v \Theta||^2$ across the 10,000 iterations of the Markov chain.
Mixing was generally good, with the median effective sample size
(the equivalent numbers of independent Monte Carlo samples)
for  $|| \v \Theta||^2$
 being 9,422.
For each simulated data set we obtained a posterior mean estimate
of $\v\Theta$. However, this estimate will generally have a rank higher than 4
as rank is not preserved under linear combinations.
For this reason,
the rank-4 least squares approximation to the posterior mean
was also computed as an alternative Bayesian point estimate of $\v\Theta$.

\begin{figure}[ht]
\centerline{\includegraphics[width=6in]{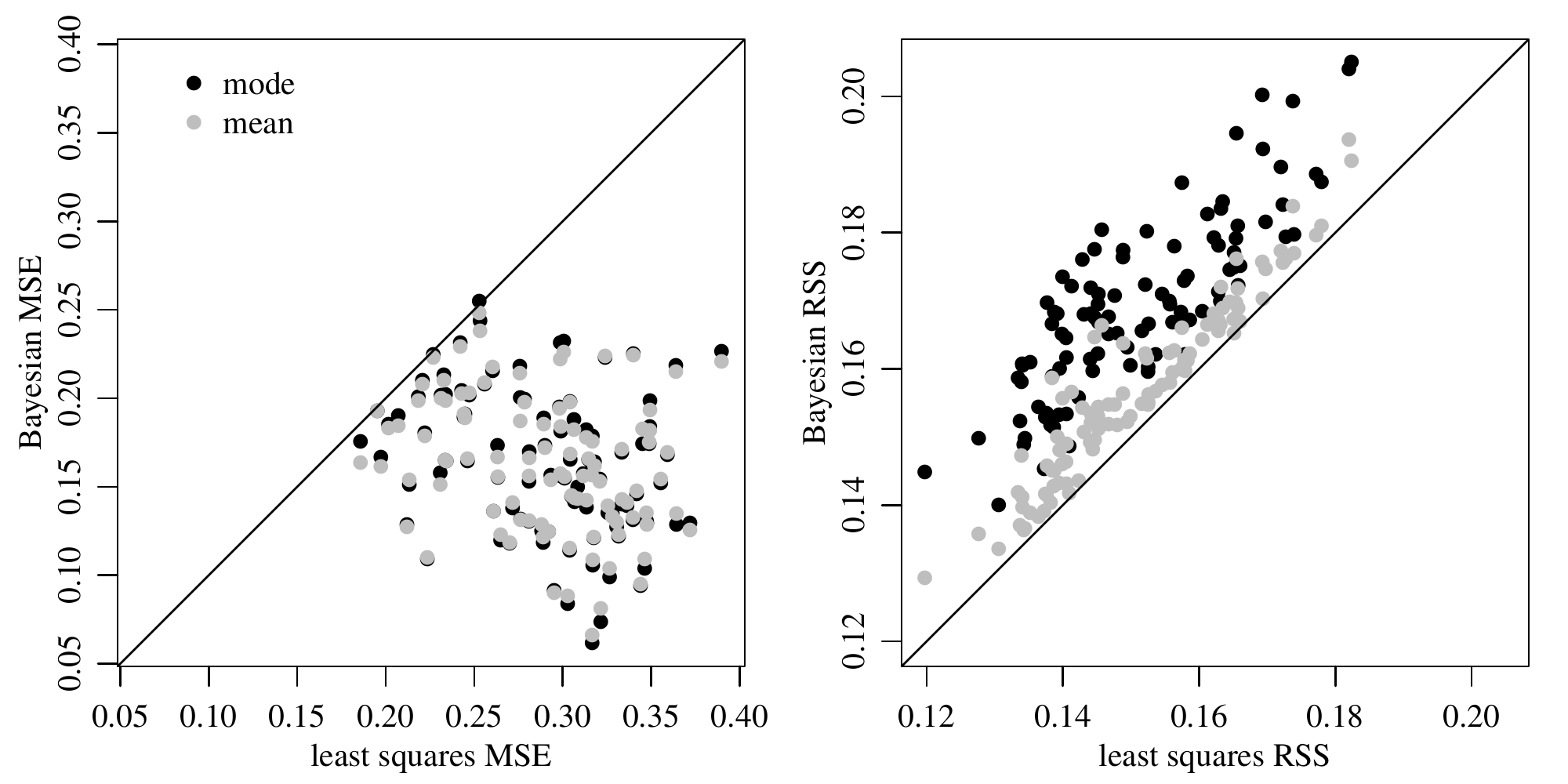}}
\caption{Comparison of MSE and RSS for different estimation methods. }
\label{fig:fig1}
\end{figure}

The results of the simulation study are summarized in Figure \ref{fig:fig1}.
For each data set and estimation method, the
ratio of $||\hat {\v \Theta} - \v \Theta||^2/
|| \v Y -  \v \Theta||^2$ was computed to assess  the performance of $\hat {\v \Theta}$
relative to the unbiased estimate $\v Y$.
In this example where the true rank of $\v \Theta$ is known, using the
reduced-rank ALS estimate is superior to using $\v Y$, giving
reductions of mean squared error of roughly 60 to 80\%.
However, the first panel of Figure \ref{fig:fig1} indicates that the
Bayesian estimators provide a substantial further reduction in MSE,
amounting to an additional reduction of 41\%
on average and up to 80\% for particular data sets.
Also, note that the rank-4 Bayesian point estimate performs essentially
the same as the posterior mean estimate, even though the latter may be
of rank higher than 4.

One possible explanation for the superiority of the Bayesian approach over
ALS is that the latter does not explore as much of the
parameter space as an
 MCMC algorithm.  The second panel of Figure \ref{fig:fig1},
which  plots
the relative residual sum of squares (RSS)
$||\m Y - \hat {\v\Theta}||^2/||\m Y||^2$ for the ALS estimate versus
the two Bayes estimates,
suggests that
this is not the case. This plot indicates that
$\hat {\v \Theta}_{\rm LS}$ is in fact closer to $\m Y$
than $\hat {\v \Theta}_{\rm HB}$  for every simulated data set. This
observation, together with the superiority of the Bayes estimate
in terms of estimating $\v \Theta$, suggests that the ALS  procedure
tends to overfit.

For each of the 100 simulated data sets an
alternative Bayesian estimate of $\v\Theta$ was also obtained,  in which
the elements $u^{(k)}_{i,r}$ of the $\v U$-matrices were  assumed to be
{\it a priori}
independent normal$(0,100)$ random variables. This non-hierarchical
approach fixes the amount of regularization, and does not recognize
patterns in $\v \Theta$ that could be represented by correlations
among the latent factors.  Not surprisingly, estimates obtained
from this approach generally had higher MSEs than the estimates based on
the hierarchical model (in 99\% of the cases using the posterior mean
estimates, and 92\% of the cases using rank-4 point estimates).

\subsection{Misspecified rank}
A more realistic data analysis situation is one in which the true rank of
$\v \Theta$ is not known. In this subsection we investigate
the MSEs of $\hat {\v \Theta}_{\rm LS}$
$\hat {\v \Theta}_{\rm HB}$
for estimating the rank-4 arrays generated as described above,
but when
the assumed rank is $R\in \{1,\ldots, 8\}$.
\begin{figure}[ht]
\centerline{\includegraphics[width=6.5in]{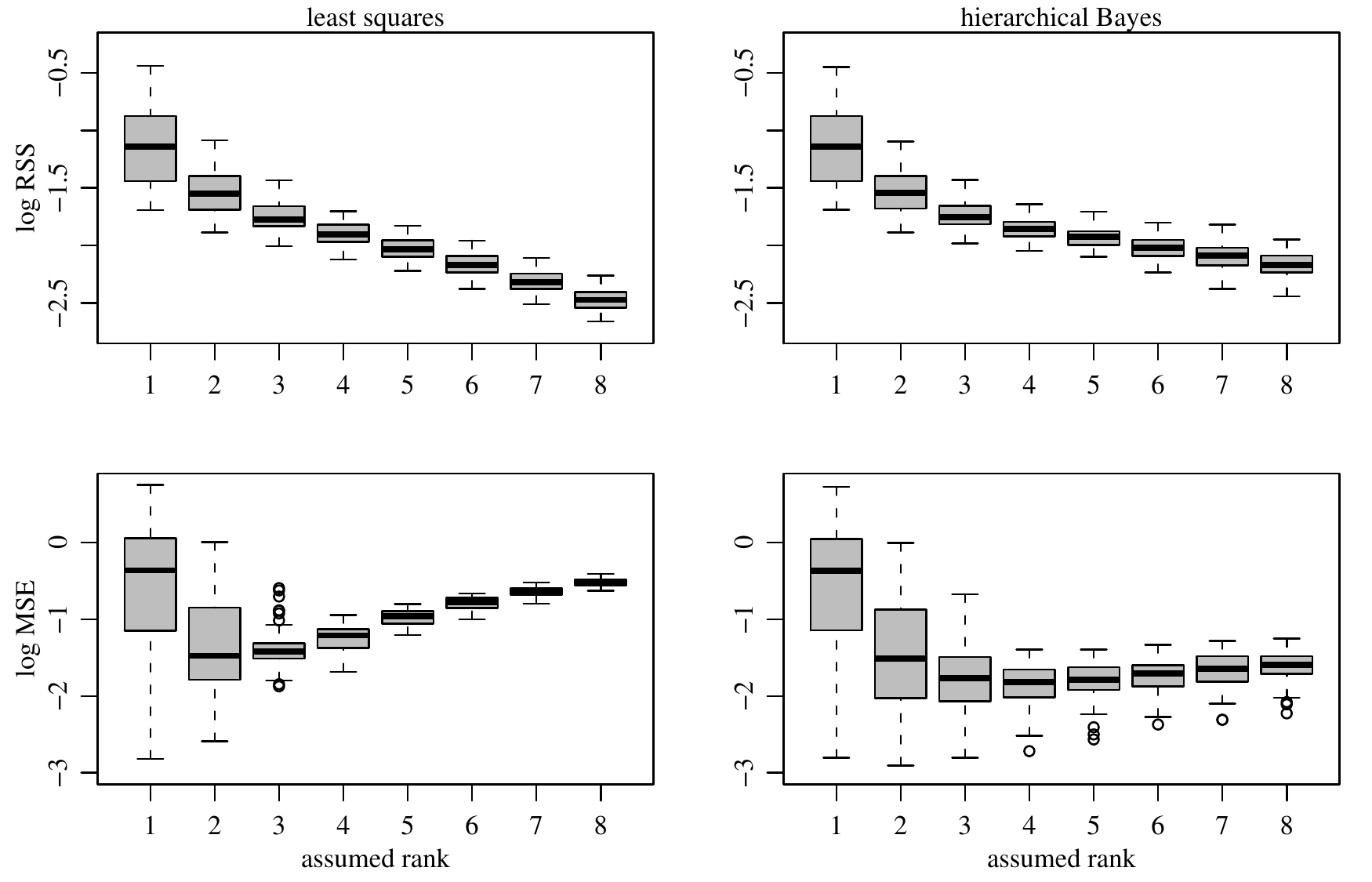}}
\caption{RSSs  and MSEs under different presumed ranks and estimation methods.}
\label{fig:fig2}
\end{figure}
Using the same simulation and estimation procedures as described
in the previous
subsection,
a  $\hat {\v \Theta}$   was obtained for
each of the 100 simulated $\v \Theta$-arrays
and for each combination of the two estimation methods and
ranks $R\in \{1,\ldots, 8\}$. For each of these $100\times 2\times 8$ estimates,
a relative
MSE $||\hat {\v \Theta} - \v \Theta||^2/
|| \v Y -  \v \Theta||^2$  and RSS
$|| \v Y - \hat {\v \Theta}||^2/ 
 || \v Y||^2$
was computed as before.
The first of these measures the fidelity of the
estimate to the true underlying
parameter, and the second to the
the data.

Summaries of the results are plotted in the four
panels of Figure \ref{fig:fig2}. For example, each
boxplot in the top row of plots
summarizes the 100 RSS values of the ALS estimates
assuming a given  rank. As expected, as the rank increases the
percentage of the variation in $\v Y$ explained by the ALS estimate
goes up and the RSS goes down.
However, the first plot of the bottom row shows that increasing the rank
of the ALS estimate beyond 3 generally increases the MSE.
In contrast, the MSE of the hierarchical estimate
 $\hat {\v \Theta}_{\rm HB}$
generally achieves a minimum at the actual rank of 4,
and increases relatively slowly as the assumed rank is increased beyond 4.
This suggests that the hierarchical Bayes approach
is more robust to overfitting than the least squares method.
Since the ``true'' rank of $\v \Theta$
is generally not known, it may be desirable to fit a model with a
moderately large rank in the hopes of capturing as much of
$\v \Theta$ as possible.
The above results suggest that a
hierarchical Bayes estimate may be preferable in such situations,
as it provides a more stable estimate of $\v \Theta$ across different
choices of the presumed rank.

\subsection{Rank selection}
We now consider the possibility of estimating the rank $R$ from the
observed data array $\m Y$. One popular model selection procedure is
to minimize the
Bayesian information criterion, or BIC \citep{schwarz_1978}.
The BIC for a given model and data set $y$ is
 $-2\ln  p(y| \hat \theta ) + p \ln n$,
where  $\hat \theta$ is the parameter estimate, $p$ is the dimension of
$\theta$ and $n$ is the sample size.
In practice, the BIC can be computed for a range of different models, and
the one giving the smallest BIC is selected.  This procedure
favors models that fit well (in terms of likelihood) but penalizes
model complexity.

As pointed out by \citet{pauler_1998}, for hierarchical models the
number of parameters can be ambiguous. As a remedy,
\citet{spiegelhalter_best_carlin_vanderlinde_2002} proposed the
deviance information criterion, or  DIC which can be computed from
output of a Markov chain.  The DIC  is given by
  $\bar D + \tilde p$ , where $\bar D$ is the average value of
$-2\ln p(y|\theta) $ across iterations  of the Markov chain, and $\tilde p$
is the ``effective number of parameters'', given by
$\tilde p = \bar D  + 2 \ln p(y|\hat \theta)$, where $\hat \theta$ is
an estimate of $\theta$.
For our model the parameters are $\v \Theta$ and $\sigma^2$,
and we take our estimates to be
the posterior mean of $\v \Theta$
and the mean residual  error under the posterior mean, respectively.

For each of the 100 simulated data sets described above we
computed the DIC for each value of $R\in \{1,\ldots, 8\}$, and took
our ``estimate'' $\hat R$ of $R$ to be the
rank for which the DIC was minimized.
%The fraction of times $\hat R$ took on the values
%$\{1,\ldots,8\}$ was $\{ 0.08, 0.15, 0.27, 0.28, 0.06, 0.07, 0.04, 0.05 \}$,
%making 
As shown in the second row of Table \ref{tab:rselect}, 
the true rank of $4$ was the most frequently selected value of $\hat R$, 
followed
closely by $3$. The fact that $\hat R=3$ was selected 27 times is somewhat
ameliorated by the fact that
in 15 of these instances the ``best'' rank in terms of MSE turned out
to be $3$ (11 cases)  or $2$ (4 cases).

To further evaluate the BIC procedure, we also reran the entire simulation
study when the true rank was $R=2$ and when it was $R=6$.
For the case of $R=2$, the DIC 
%selection fractions were
%$\{ 0.10, 0.74, 0.07, 0.05, 0.02, 0.01, 0.01 \}$, 
selected $\hat R=2$ in 74\% of the cases, 
indicating that
in this situation 
the true rank can be identified with a high degree of accuracy.
Rank selection with
DIC was more problematic when the true rank was 6. 
%for which the selection proportions were
%$\{0.07, 0.18, 0.19, 0.17, 0.10, 0.08, 0.09 ,0.12\}$.
As we would hope, the
distribution of ranks selected here is somewhat shifted to the right from the
distribution of selected ranks when $R=4$, but as indicated in the table, 
the true rank of 6
can not be identified accurately with DIC.
However, the DIC is not as bad
in terms of obtaining the rank that gives the best approximation
to the true $\v \Theta$ in terms of MSE.
For example,
75\%  of the 71 simulated data sets for which $\hat R$ was less than 6
also attained their minimum MSE at an $R$-value less than 6.
In particular, the seven data sets for which $\hat R=1$ also attained
their minimum MSE with a rank 1 model.

\begin{table}
\begin{center}
\begin{tabular}{c|cccccccc}  
 \multicolumn{1}{c}{ }&  \multicolumn{8}{c}{ $\hat R$ } \\  
 $R$     &   1   &  2   &  3   &  4   &  5   &  6    &  7  &  8  \\ \hline
 2    &  0.10 & 0.74 & 0.07 & 0.05 & 0.02 &  0.01 & 0.01  & 0.00 \\
 4    &  0.08 & 0.15 & 0.27 & 0.28 & 0.06 &  0.07 & 0.04  & 0.05 \\
 6    &  0.07 & 0.18 & 0.19 & 0.17 & 0.10 &  0.08 & 0.09  & 0.12
\end{tabular}
\caption{Rank selection using DIC: $R$ is the rank under which 100 simulated data sets were generated. Entries in the table give the 
 percentage of simulations for which $\hat R$, the DIC-optimal rank, took on the values 1 through 8.}
\end{center}
\label{tab:rselect}
\end{table}

\section{Example: Multiway means for cross-classified data }
Large scale surveys collect data on a variety of numerical and 
categorical variables. 
Numerical data are often summarized by computing
sample averages for combinations of a set of categorical variables.
For example, letting
$\v y$ be a $p$-dimensional vector of numerical variables  and
$\v x $ a $K$-dimensional vector of categorical variables, interest may
lie in the population average
of $\v y$ for a given value of $\v x$, which is denoted as
$\v \mu_{\rm x}\in \mathbb R^p$.
However, if the number of categorical variables or
their number of levels is large compared to the sample size, then we may lack
sufficient data
to provide stable estimates for each $\v \mu_{\rm x}$ separately. 
For example, 
the 2008 General Social Survey includes data
on the following six variables:
\begin{itemize}
\item  $y_1$ ({\tt words}): number of correct answers out of 10 on a vocabulary test; 
\item $y_2$ ({\tt tv}): hours of television watched in a typical day;
\item $x_1$ ({\tt deg}) highest degree obtained: none, high school, Bachelor's, graduate;
\item $x_2$ ({\tt age}): 18-34, 35-47, 48-60, 61 and older;
\item $x_3$ ({\tt sex}): male or female;
\item $x_4$ ({\tt child}) number of children: 0, 1, 2,  3 or more.
\end{itemize}
Complete data for these variables are available for 1116 
survey participants.  However, there are $4\times 4\times 2 \times 4 =128$ 
levels of $\v x$. More than half of these cells have 5 or fewer observations 
in them, and about 75\% have less than 12 observations. 
As such, an estimator of  $\v \mu_{\rm x}$ that uses only data from group $\v x$, that is
$\{ \v y_i : \v x_i = \v x \}$, will be subject to a large sampling variance. 

\subsection{A multilinear model for group means}
Statistical remedies to this problem typically allow the estimate 
of $\v \mu_{\rm x}$ to depend on data from groups other than that corresponding 
to $\v x$. One  such approach is to
parameterize the set of multivariate 
means $\{\v \mu_{\rm x} : \v x\in \mathbb X\}$ by a smaller 
number of parameters. Another approach is via  a hierarchical model that 
 allows for 
the shrinkage of set of parameters towards a common group center. 
Here we consider the following model which has both of these features:
\begin{eqnarray}
\{\v y_{i} : \v x_{i} = \v x \} &\stackrel{\rm iid}{\sim} &
\mbox{multivariate normal}(\v \mu_{\rm x} ,\v \Sigma)   \label{eq:hm1} \\
\v \mu_{\rm x} & =& \v \beta_{\rm x} + \v \gamma_{\rm x}  \label{eq:hm2} \\
\{ \v \gamma_{\rm x} : \v x\in \mathcal X  \}  &\stackrel{\rm iid}{\sim} & 
   \mbox{multivariate normal}(\v 0 ,\v \Omega)  \label{eq:hm3}. 
\end{eqnarray}
Equation \ref{eq:hm1} indicates that the data within a cell 
are modeled as multivariate normal, with cell-specific means and a common covariance 
matrix. Equations \ref{eq:hm2} and \ref{eq:hm3}  express each  $\v\mu_{\rm x}$ 
as equal to a ``systematic'' component $\v \beta_{\rm x}$ plus 
patternless noise $\v\gamma_{\rm x}$. 
\begin{figure}[ht]
\centerline{\includegraphics[width=6.5in]{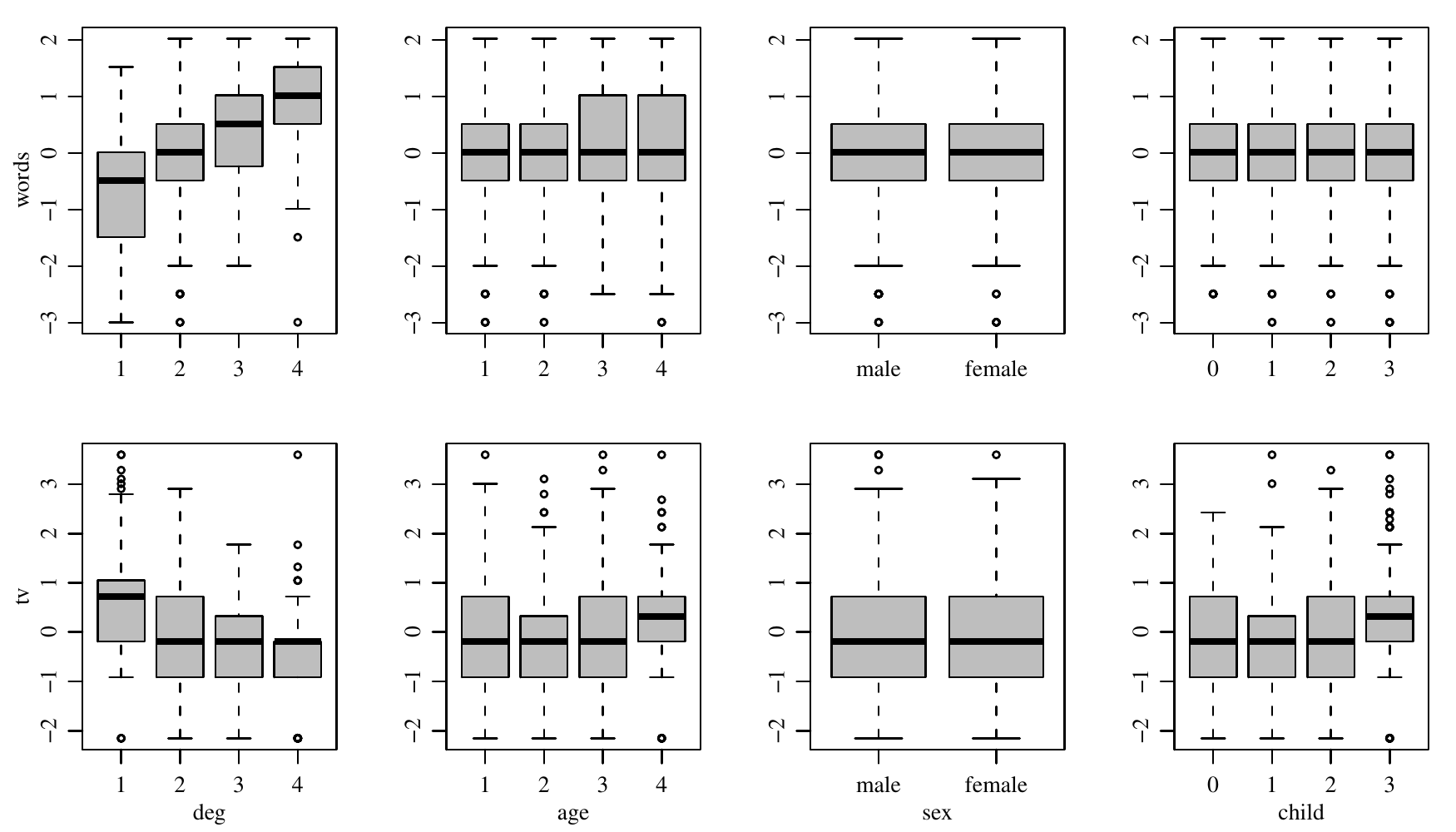}}
\caption{Marginal distributions of vocabulary score and television hours watched
for different levels of degree, age, sex and
number of children.}
\label{fig:wtv_data}
\end{figure}
The collection $\{ \v \beta_{\rm x} : \v x \in \mathcal X \}$ can be
represented as an $m_1 \times   \cdots \times m_K 
\times p$ array $\m B$, where 
$m_k$  is the number of levels of categorical variable $x_k$. 
These values are not separately estimable from the noise $\v \gamma_{\rm x}$
unless we assume $\m B$ lies in a restricted subset of the set of 
arrays of this size, such as the set of rank-$R$ arrays. 
In this setting, where one of the modes of the array represents variables and 
 each other mode represents the different levels of a single categorical 
variable, it is useful to express the array decomposition as follows:
 \begin{eqnarray*}
\m B & =& \langle \m U^{(1)} , \ldots,  \m U^{(K)} , \m V \rangle  \  , \ \mbox{ or equivalently} \\
\v \beta_x &=& \m V ( \m u^{(1)}_{x_1} \circ \cdots \circ \m u^{(K)}_{x_K}  ). 
\end{eqnarray*}
The equations above describe  a hierarchical model in which  the 
heterogeneity among $\{ \v \mu_{\rm x} : \v x \in \mathcal X\}$ is 
centered around a low-dimensional array
 $\m B = \{ \v \beta_{\rm x} : \v x \in \mathcal X\}$.
Such a model is similar to representing an interaction term
in an ANOVA  with 
a reduced rank matrix \citep{tukey_1949,boik_1986,boik_1989}. However, 
the hierarchical approach used here 
allows 
for consistent estimation of each $\v \mu_{\rm x}$, 
but shrinks 
towards the lower-dimensional representation $\m B$ when data are limited.

Estimation for this model can proceed as described in Section 4 with a few 
modifications. 
As before, a Gibbs sampler can be used to approximate 
the posterior distribution of the unknown parameters. 
Using a conjugate inverse-Wishart prior 
distribution for $\v \Sigma$ and the other prior distributions as 
in Section 4, 
 one iteration of the Markov chain 
is as follows:
\begin{enumerate}
\item  sample $\v \Sigma\sim p(\v \Sigma| \{ \v y_i:i=1,\ldots, n\},\{\v \mu_{\rm x} :\v x\in \mathcal X\} )$, 
 an inverse-Wishart distribution; 
\item sample  
$\v\mu_{\rm x} \sim p( \v\mu_{\rm x} | \{ \v y_i: \m x_i=\v x\}, \v\beta_{\rm x}, \v \Sigma)$, a multivariate normal distribution for each $\v x\in \mathcal X$; 
\item sample $\v \Omega \sim p(\v \Omega| \{\v \mu_{\rm x},\v\beta_{\rm x}:\v x\in \mathcal X\},\m V)$, an inverse-Wishart distribution;  
\item iteratively sample $\{ \m U^{(k)},k=1,\ldots, K\}$ 
as in 
Section 3; 
\item sample $\m V \sim p(\m V|\m U, \{ \v \mu_{\rm x} : \v x\in \mathcal X \}, \v \Omega)$, 
a matrix normal distribution. 
\end{enumerate}
Derivations of the full conditional distributions are straightforward and 
are available from the author and in the computer code available at the 
author's website. 
Provided here are a few comments 
that describe some of the calculations: 
Let the model for the $p\times R$ matrix $\v V$ be such 
that the $R$ columns are i.i.d.\ multivariate normal with a zero mean vector 
and covariance equal to $\v \Omega$. Doing so links the scale of 
the factor effects for $\v \mu_{\rm x}$ to the scale of the 
across-group differences $\v \gamma_{\bf x}$. 
Writing $\tilde {\v \mu}_{\rm x} = \v \Omega^{-1/2} \v \mu_{\rm x}$ and
  $\tilde {\m V} = \v \Omega^{-1/2}\v V$, 
we have
\begin{eqnarray*}
\tilde{\v \mu}_{\bf x} &=& \tilde{\m V}  ( \v u^{(1)}_{x_1} \circ \cdots \circ
  \v  u^{(K)}_{x_K} ) + \tilde{\v\gamma}_{\bf x}  \ , \ \mbox{ with } \\
\{ \tilde{\v \gamma}_{\bf x} \} &\stackrel{\rm iid}{\sim}  &
\mbox{multivariate normal}(\v 0 ,\m I).
\end{eqnarray*}
From this, we see that sampling from the full conditional distribution of 
$\v U^{(1)},\ldots, \v U^{(K)}$ can be done just as in Section 3.2, with 
$\sigma^2$ replaced by 1 and the observed array data replaced by the 
 values of the array defined by  $\{ \tilde {\v \mu}_{\rm x} :
   \v x\in \mathcal X \}$. Similarly, the full conditional 
of $\tilde {\v V}$ is the matrix normal distribution from Section 3.1, 
again with $\sigma^2$ replaced by 1 and 
$\{ \tilde {\v \mu}_{\rm x} :
   \v x\in \mathcal X \}$ taking the place of the observed array data. 
A value of $\m V$ can be generated from its full conditional distribution 
by sampling $\tilde {\v V}$ from this matrix normal distribution and then 
setting $\m V= \v \Omega^{1/2} \tilde {\v V}$. 
Finally,  note that the inverse-Wishart full conditional distribution of 
for $\v \Omega$ depends on $\m V$. If we have 
$\v \Omega \sim$ inverse-Wishart$(\v \Omega_0^{-1},\eta_0)$ 
then the full conditional
distribution of $\v\Omega$ is inverse-Wishart$(\v \Omega_1^{-1}, \eta_1)$
where
$\eta_1 = \eta_0 + R + \prod_{k=1}^K m_k  $ and 
$\v \Omega_1 = \v \Omega_0 + \m V^T\m V +  \sum_{\bf x} (\v \mu_{\rm x} - \v \beta_{\rm x} ) 
   (\v \mu_{\rm x} - \v \beta_{\rm x} )^T$. 

\begin{figure}[ht]
\centerline{\includegraphics[width=6.5in]{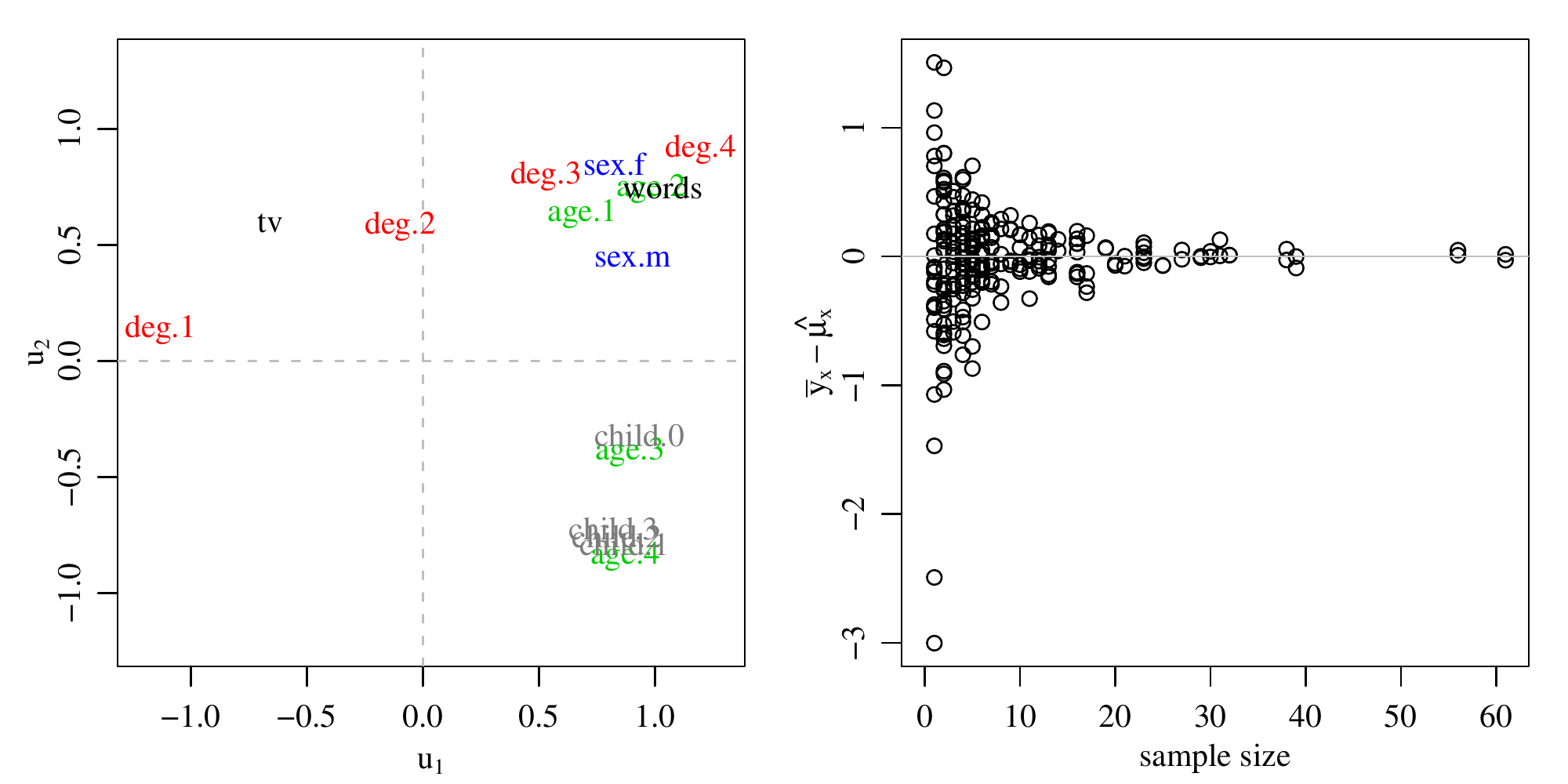}}
\caption{Posterior estimates of factor scores, along with the amount 
of shrinkage as a function of cell-specific sample size. }
\label{fig:wtv_post}
\end{figure}

\subsection{Posterior analysis of GSS data}
We now discuss posterior inference for the GSS data based on the above 
model and estimation scheme. 
The  numerical variables $y_1$ ({\tt words}) and $y_2$  ({\tt tv})
were first centered and scaled to have 
zero mean and unit variance. 
Prior distributions for the covariance 
matrices $\v \Sigma$ 
and $\v\Omega$ were taken to be independent inverse-Wishart 
distributions
with  $p+1=3$ degrees of freedom each and centered around the sample 
covariance (correlation) 
matrix of $\{ y_{i,1},y_{i,2},i=1,\ldots, n\}$. Doing so  gives these
prior distributions
an empirical basis  while still keeping them relatively weak. Such priors are 
 similar 
to the ``unit information'' prior distributions described in 
\citet{kass_wasserman_1995}. 
A rank-2 model for the array of means was used so that the estimated 
factor effects could represented with a simple two-dimensional plot. 

The algorithm described above was used to construct a Markov chain consisting of 22,000 iterations, the 
first 
2,000  of which
 were discarded to allow for convergence to the stationary distribution. 
Parameter values were saved every 10th iteration, leaving 2,000 saved values 
for Monte Carlo approximation. Mixing of the Markov chain was examined 
by inspecting the sequences  of saved values 
of $\v \Sigma$, $\v \Omega$ and the average value of 
$\{\v \beta_{\rm x}\}$ across levels of $\v x$.
The effective sample sizes  for these parameters
were all over 1,000. 
Some summary descriptions of the resulting posterior  estimates 
are shown in Figure \ref{fig:wtv_post}. The first panel plots point 
estimates of the latent factors $\m V$ and $\m U^{(1)},\ldots, \m U^{(4)}$. These were obtained 
as follows: A posterior mean array 
$\bar {\m B}$
 was obtained from the 2,000 saved values of $\m B$ from  the Markov 
Chain. This array is not quite a rank-2 array, as rank is not generally 
preserved 
under array addition. An alternating least-squares algorithm was 
performed on $\bar {\m B}$ to obtain 
a rank-2 point estimate  $\hat {\m B}$ and a multiplicative decomposition 
in terms of matrices $\hat {\m V}$, $\hat {\m U}^{(1)},\ldots, \hat{\m U}^{(4)}$.
The difference between $\bar {\m B}$ and  $\hat {\m B}$ was small, with 
 $|| \bar {\m B} -\hat {\m B} ||^2/||\bar {\m B}||^2 =0.00011$.
These point estimates of the latent factors
are shown in the first panel in Figure \ref{fig:wtv_post}. For example, 
the matrix $\hat {\m U}^{(1)}$ represents the multiplicative effects of 
{\tt deg}, and consists of a two-dimensional vector for each level of 
this variable. These vectors are plotted in the figure with ``deg.1'' representing 
no degree, ``deg.2'' a high school degree, and so on. Similarly, 
the matrix $\hat {\m V}$ has a two-dimensional vector for each of the 
two numerical  variables. To interpret the figure, note that the estimated 
mean for either numeric variable in any cell can be obtained by 
coordinate-wise multiplication and then addition 
of the latent factor vectors. For example, the proximity of the 
``words'' vector to the ``deg.3'' and ``deg.4'' vectors indicates that these 
two groups have higher mean vocabulary scores  than the other two degree
categories. Similarly, the close proximity of the ``child.1'', ``child.2''
and ``child.3'' vectors indicates lack of heterogeneity in the means for 
three of these four categories 
across levels of the 
other $\v x$-variables. 
Finally, note that some care should go into interpreting the figure, 
as the array $\m B = \langle \m U^{(1)},\ldots, \m U^{(4)}, \m V\rangle$ is invariant to certain transformations of the factors. 
For example, multiplying either the first or second column of 
each of 
an even number of factor matrices by -1 does not change the value of $\m B$. 

The second plot in Figure \ref{fig:wtv_post} highlights how the 
estimated  cell means $\{ \hat{\v\mu}_{\rm x}\}$ differ from the
empirical cell means $\{  \bar {\v y}_{\rm x}\}$
as a function of sample size. 
This plot indicates what we would 
 expect from a hierarchical model: The difference between 
estimated cell mean and empirical cell mean decreases with increasing sample 
size. A cell with a large sample size will have 
$\bar {\v y}_{\rm x} \approx\hat{\v \mu}_{\rm x}$ , 
whereas a cell with  a small sample size will have an estimated mean
 $\hat{\v\mu}_{\rm x}$ shrunk towards the reduced-rank value 
$\hat{\v \beta}_{\rm x}$. 
Note that without the multiplicative effects
in  Equation \ref{eq:hm2} of the hierarchical model, the cell means would 
all be shrunk towards a common vector, regardless of the value of $\v x$. 
In contrast, the hierarchical multiplicative effects model allows cell-specific shrinkage, as estimated
by the reduced rank array $\hat{\m B}$.

An alternative  approach to the analysis of these data might
involve MANOVA or a  hierarchical model similar to the one above but in which
$\v\beta_{\rm x}$ is parameterized  in terms of additive effects, so that
 $\v\beta_{\rm x} = \v u_{x_1}^{(1)} + \cdots + \v u_{x_K}^{(K)}$
with each $\v u_{x_k}^{(k)} \in\mathbb R^p$. Such additive models
have representations as  multilinear models, although of course
they are restricted to be additive.
For comparison,  an additive MANOVA model was fit and
 the average value of $( \bar {\v y}_{\rm x}  - \hat {\v \beta}_{\rm x})
^
2$ was computed,  measuring the lack-of-fit of the additive model. 
This value was about the same  as
the corresponding value for the multilinear model for the {\tt tvhours}
variable, but 15\% larger for the {\tt words} variable.
This indicates that some patterns among the cell means for 
{\tt words} cannot be represented with an additive model. 
In general, we may expect that some aspects of the heterogeneity among the
$\v \mu_{\bf x}$'s will not be additive. In such situations, 
it may be
preferable to use a multiplicative model whose complexity can be
controlled with the choice of the rank $R$
rather than to have to consider 
the inclusion and estimation of a variety of higher-order interaction terms.

\section{Example: Analysis of longitudinal conflict data}
The theory of the Kantian peace holds that militarized interstate 
disputes are less likely to occur between democratic countries. 
\citet{ward_siverson_cao_2007} evaluate this theory using 
international cooperation and conflict data from the cold 
war period. The data include records of militarized  conflict 
and cooperation every five years from 1950 to 1985, along with 
economic and political characteristics of the countries. 
In this section we analyze a subset of the data from 
\citet{ward_siverson_cao_2007}. 
These data include cooperation, conflict and gross domestic product data (gdp)
for each of $m=66$ countries every fifth 
year, $t\in \{1950,1955,\ldots, 1980,1985\}$.
Additionally, each country in each  of these years has 
a polity score, measuring 
the level of openness in government. A positive polity score is given 
to democratic states, while a negative score is given to 
authoritarian states. 

The cooperation and conflict data  form a three-way array with two 
modes representing country pairs and one mode representing time. In this 
section we will fit an ordered probit model of 
cooperation and conflict data as a function of   
gdp and polity. Specifically, for each unordered pair $\{i,j\}$ 
of countries and 
each time $t$, our data are as follows:
\begin{itemize}
\item[] $y_{i,j,t} \in \{ -5,-4,\ldots, +1,+2\}$,
indicating the level of military cooperation
(positive) or conflict (negative) between countries $i$ and $j$ in year $t$; 
\item[] $x_{i,j,t,1}= \log {\rm gdp}_i + \log {\rm gdp}_j $, the sum 
of the log gdps of the two countries; 
\item[] $x_{i,j,t,2}=  (\log {\rm gdp}_i )\times (\log {\rm gdp}_j)$, 
the product of the log gdps;
\item[] $x_{i,j,t,3}= {\rm polity}_i \times {\rm polity}_j$, where ${\rm polity}_i \in \{-1,0,+1\}$; 
\item[] $x_{i,j,t,4}= ({\rm polity}_i>0) \times ({\rm polity}_j>0)$.
\end{itemize}
The sample space for $y_{i,j,t}$ is ordered but the 
scale is not meaningful:
 The difference between $y=0$ and $y=1$ 
is not comparable to the difference between $y=-5$ and $y=-4$. For this 
reason we use the following 
ordered probit model to relate $y_{i,j,t}$ to $\v x_{i,j,t}$:
\begin{eqnarray*} 
z_{i,j,t} &=& \v \beta^T \v x_{i,j,t} + \gamma_{i,j,t}  \\
y_{i,j,t}& = & \max \{ k : z_{i,j,t}> c_k ,  k\in \{-5,-4,\ldots, +1,+2\} \}. 
\end{eqnarray*}
In this model the parameters to estimate include the regression 
coefficients $\v \beta$ and the cutoffs $\v c=(c_{-4},\ldots, c_{+2})$, 
with $c_{-5}= - \infty$. 
The usual probit regression model would assume
the $\gamma_{i,j,t}$'s are  independent standard normal variables 
(standard, as the scale of these error terms is 
not separately identifiable from 
$\v \beta$ and $\v c$). 
However, results of  \citet{ward_siverson_cao_2007}
 suggest that the residuals from 
regression models of international relations data are generally not 
patternless. For example, we might expect $\gamma_{i,1,t} ,\ldots, 
  \gamma_{i,66,t}$ to 
exhibit statistical correlation, as these residuals are all associated with 
country $i$. More subtle might be higher order patterns common in relational 
data:  If $i$ and $j$ have a positive 
relationship and $j$ and $k$ have a positive relationship, then a 
positive relationship between  $i$ and $k$  is more likely.

\citet{hoff_2008a} describes how two-way factor models can be used to 
represent patterns in ordinal
matrix-valued relational and social network data. 
Here we extend this idea, using a three-way factor model to 
represent the longitudinal relational patterns 
represented by the array  $\v \Gamma =\{\gamma_{i,j,t}\}$. 
Specifically, the following factor model is proposed:
\begin{eqnarray*}
\gamma_{i,j,t} &=& \langle \v u_i ,\v u_j, \v v_t \rangle + \epsilon_{i,j,t} \ , \ \mbox{with} \\
\{ \epsilon_{i,j,t} =\epsilon_{j,i,t} \} &\stackrel{\rm iid}{\sim} &  
{\rm normal}(0,1).
\end{eqnarray*}
The $\v u_j$'s are vectors representing heterogeneity among the 
countries and the $\m v_t$'s represent heterogeneity over time. 
This is a modification of the usual three-way PARAFAC representation to accommodate 
the fact that the data are symmetric $(y_{i,j,t}=y_{j,i,t})$. 
This model has a simple interpretation:
Letting $\v \Gamma_t=\{\gamma_{i,j,t}: (i,j) \in \{1,\ldots,m\}^2 \}$, 
we have 
\begin{eqnarray*} 
\v \Gamma_{t} &=&  \m U \v \Lambda_t \m U^T + \m E_t \ , \ \mbox{where}  \
\m U = ( \m u_1,\ldots, \m u_m )^T  \ \mbox{and}  \ \v \Lambda_t = {\rm diag}( \v v_t ).
\end{eqnarray*}
This symmetric version of the PARAFAC model is analogous to 
a type of  eigenvalue decomposition of  the collection of square matrices 
$\{ \v \Gamma_{1950},\ldots, \v \Gamma_{1985} \}$ in which the 
eigenvectors are held constant across matrices, but the eigenvalues are 
allowed to vary. The resulting matrices $\m U \v \Lambda_t \m U^T$ are 
then each symmetric and of rank $R$. Heterogeneity across countries 
is determined by the rows of $\m U$, and 
heterogeneity across time is determined by the $\v\Lambda_{t}$'s.

\begin{figure}[ht]
\centerline{\includegraphics[width=6in]{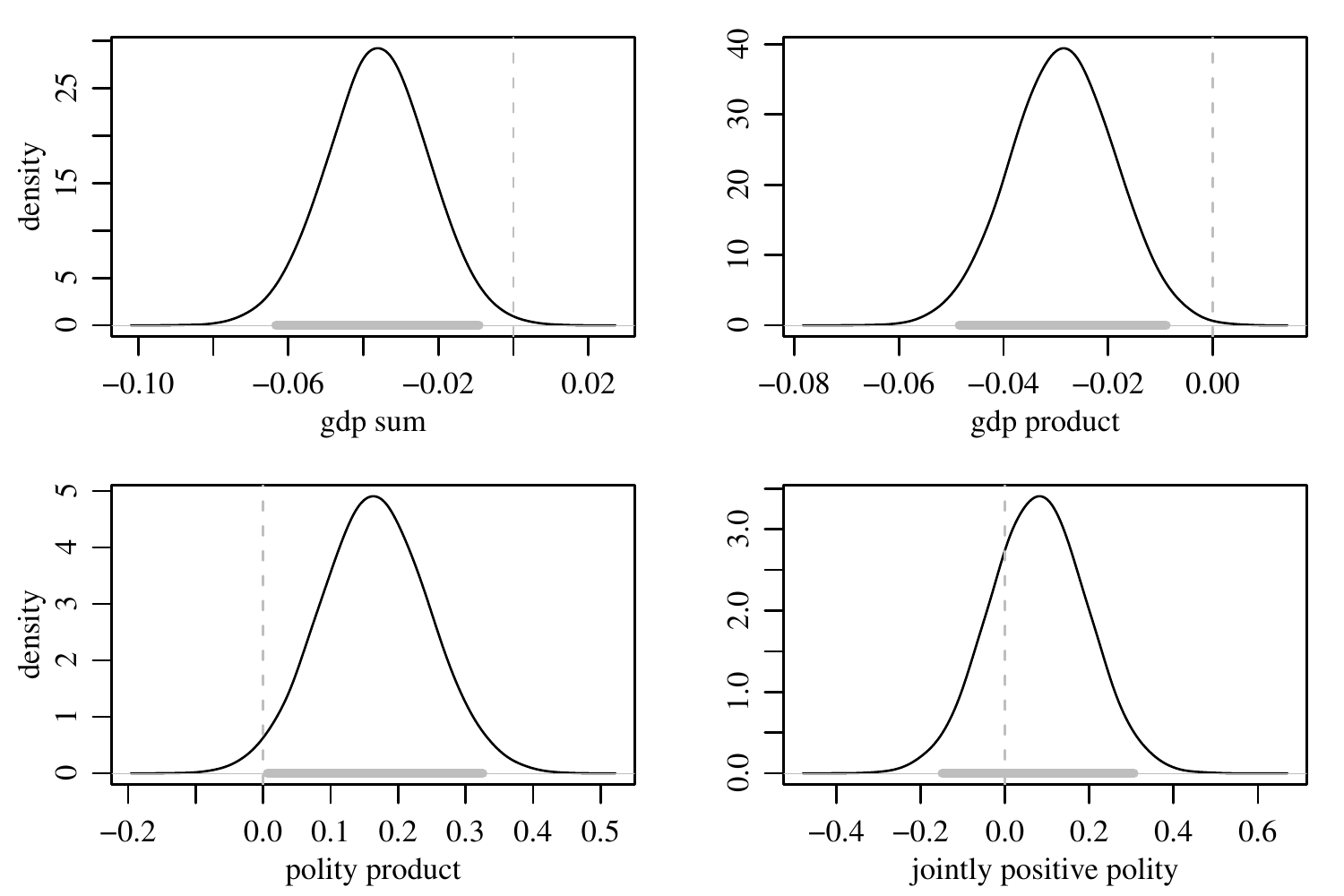}}
\caption{ Posterior densities for the elements of $\v \beta$.
  Gray lines are 95\% confidence intervals. }
\label{fig:betapost}
\end{figure}

\begin{figure}[ht]
\centerline{\includegraphics[width=7in]{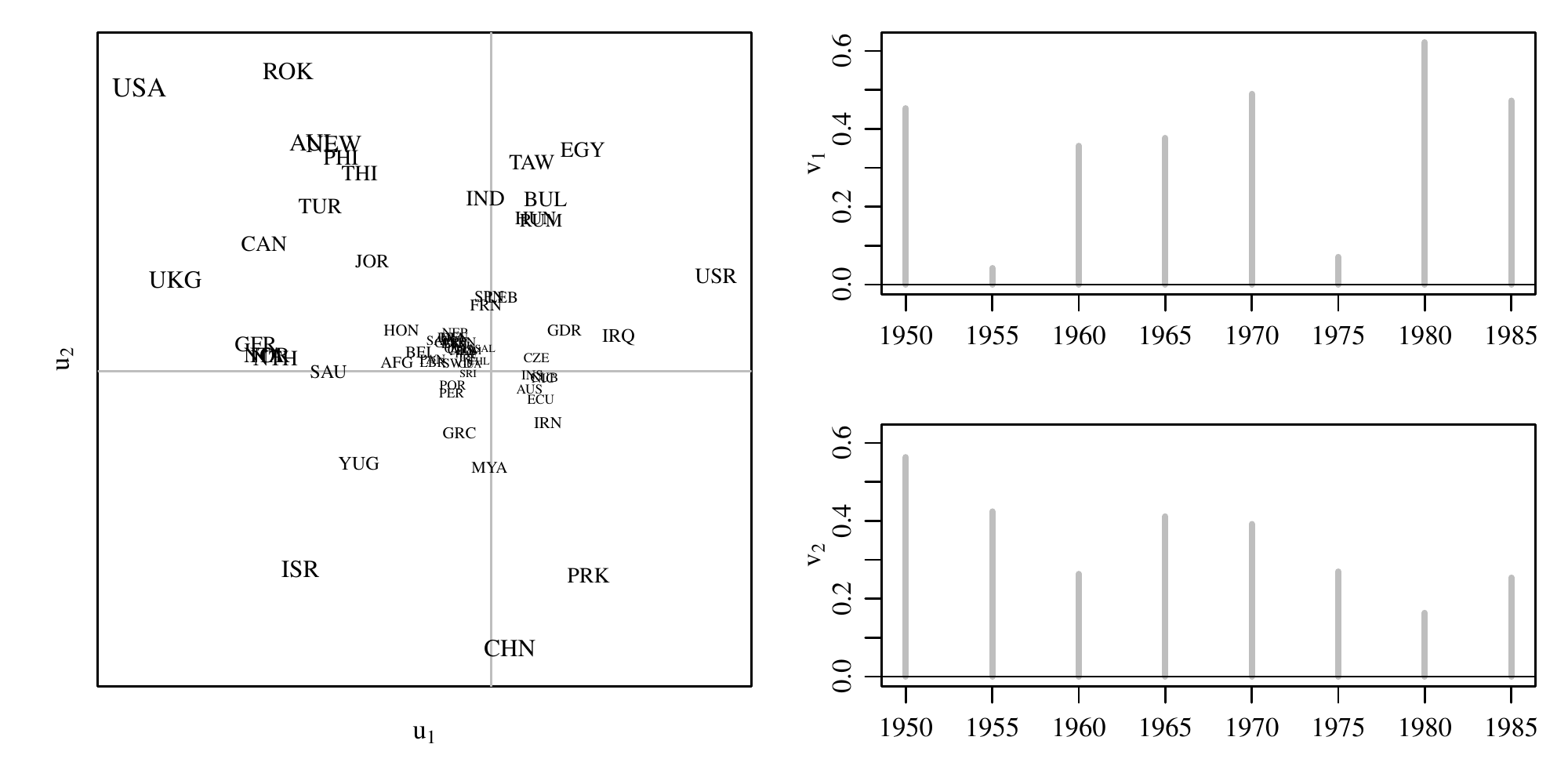}}
\caption{Posterior estimates of the country- and time-specific factors. }
\label{fig:uwpost}
\end{figure}

The unobserved quantities in this model include the latent variable 
array $\m Z$ as well as the parameters $\m U, \m V$ and $\v \beta$. 
Using the same hierarchical 
prior distributions for $\m U$ and $\m V$ described in 
Section 3.2 and a diffuse multivariate normal$(\m 0, 100\times \m I)$ 
prior distribution for $\v \beta$, we can implement a Gibbs sampler 
to  approximate the joint posterior 
distribution $p(\m Z, \m U, \m V, \v \beta | \m Y, \m X)$.
All full conditionals are standard, and are available from the 
supplementary material at the author's website. 
Using a rank-2 model, 
the Gibbs sampler was run for 505,000 iterations, dropping the first 
5,000 to allow for burn-in
and then saving the parameter values every 10th iteration. 
Convergence of the Markov  chain was monitored 
 via the sampled values of $\v \beta$. 
The effective sample sizes for the 
four regression coefficients based on the 50,000 saved scans 
were  12,548, 16,622, 1,386 and 8,878 respectively.

The plots in Figure \ref{fig:betapost} show the marginal 
posterior distributions of the four regression coefficients, along 
with 95\% highest posterior density confidence intervals. 
The results indicate a negative association between gdp and 
 the latent variable $z$, 
reflecting the fact that 
a majority of the conflicts over the cold war period involved 
economically large 
countries.
The plots in the second row indicate that  
$z_{i,j}$ tends to be larger if both $i$ and $j$ have polity scores of the 
same sign, 
but that there is not strong evidence for a further increase if 
the polities of $i$ and $j$ are both positive. 

Figure \ref{fig:uwpost} displays a summary of the posterior distribution 
of $\m U$ and 
$\m V$. This summary was obtained as follows: First, 
a Monte Carlo approximation $\hat {\v  \Theta}$ of the 
 posterior mean 
of the three-way array 
$\v \Theta =  \langle \m U , \m U, \m V \rangle $ was obtained using the 
values generated from the Markov chain. 
The alternating least-squares algorithm was then applied to 
$\hat {\v \Theta}$ to obtain values $\hat {\m U}$ and $\hat {\m V}$. 
The columns of $\hat {\m U}$ were normalized to be unit vectors, 
and the columns of $\hat {\m V}$ were then rescaled accordingly. 
The columns of the latent factor matrices  were then permuted so that
 the magnitude of the columns of $\hat {\m V}$
were in decreasing order. 
The resulting values are plotted in Figure \ref{fig:uwpost}. 
The large square plot shows the estimates of the 
two-dimensional latent factor vectors  $\{ \hat {\v u}_i \}$ for each country, 
with a larger font used for those countries with larger vectors. 
 The second column gives the values of 
$\hat v_{t,k}$, sorted chronologically. 
Since all of these values are positive, two latent vectors 
 $\{\hat {\v u}_{i_1} , \hat {\v u}_{i_2} \} $ being in similar directions 
indicates a tendency for countries $i_1$ and $i_2$ to cooperate militarily,
 whereas vectors in opposite directions indicate a tendency for conflict. 
For example, the vectors corresponding to USA and South Korea (ROK) are similar to 
each other and in the opposite direction of China (CHN) and North Korea (PRK). 
The heterogeneity of the $\hat {\v v}_t$'s over time allows for different 
patterns of conflict across the years. For example, cooperation and conflict
 in 1980 and 1985
are described primarily by the first dimension of the factors ($u_1$), whereas
events in 1955 and 1975 primary by the second ($u_2$).

\section{Discussion}
This article has presented a hierarchical version of a 
reduced-rank multilinear model 
for array data and a Bayesian method for parameter estimation. 
Unlike least-squares estimation, a Bayesian approach allows for 
regularized estimates of the potentially large number of parameters in a 
multilinear model. Unlike a non-hierarchical Bayesian approach, the 
hierarchical approach provides a data-driven 
method of regularization, and a more flexible representation of the 
patterns in the  data  array. Additionally, in a simulation study the 
estimates provided by the  
hierarchical approach showed robustness to rank misspecification, as compared 
those obtained from a least-squares or non-hierarchical approach. 

Another advantage of the Bayesian approach is that it allows for the 
incorporation of multilinear structure into a broad class of 
statistical models. For example, a least-squares approach would be 
inappropriate for the ordinal cooperation and conflict data in Section 
6, but Bayesian estimation for these data, using a probit model with multilinear effects,
is relatively straightforward. As another example, the survey 
data presented in Section 5
was not in the form of an array, 
but the cell means corresponding to the 128 levels of the 4 categorical 
variables can be represented as such. A reduced-rank multilinear model 
provides a parsimonious representation of the cell means, 
but also is more flexible than a simple additive effects model. 

An important line of future research is the study of the theoretical 
properties of hierarchical Bayesian approaches to parameter estimation 
for multiway data arrays. For a matrix model in which  
$\m Y = \v \Theta +\m E$ and $\m E$ is a matrix of normally-distributed noise, 
\citet{tsukuma_2008,tsukuma_2009}  studies Bayesian and 
hierarchical Bayesian approaches to providing admissible and minimax estimates of $\v \Theta$. One aspect of this work shows that 
under certain prior distributions on 
the singular vectors of $\v\Theta$, the Bayes estimates are equivariant 
and can be obtained by 
shrinking the singular values 
of $\m Y$. 
Such estimates are somewhat analogous to those presented in this 
article for multiway data, as shrinking the singular values of a matrix 
is similar to regularizing  the variance  of a set of multiplicative factors. 
The author is currently investigating the extent to which such similarities between the matrix and array models lead to similar 
theoretical properties of Bayesian estimates in the two cases. 
Additionally, hierarchical Bayesian procedures, like the one in this article,
often produce estimates similar to those from empirical Bayes 
 and James-Stein procedures, which have been shown to outperform 
the least-squares criterion in a variety of multivariate estimation problems
\citep{james_stein_1961,efron_morris_1973}. 
It seems likely that estimators from 
such shrinkage procedures will enjoy similar 
advantages over least squares estimation as the hierarchical model 
presented in this article. 

A popular alternative approach to shrinkage estimation for high-dimensional models is 
based on $L_1$ penalization \citep{tibshirani_1997}, in which an
estimate is obtained by minimizing the residual sum of squares plus an
$L_1$ penalty on the parameter values. In the context of estimating 
a three-way array $\v \Theta$, this could mean obtaining the value 
$\v \Theta$ that minimizes 
$|| \m Y - \v \Theta ||^2 +  \lambda \sum_{i,j,k} |\theta_{i,j,k}|$. 
However, the multiplicative parameterization of $\theta_{i,j,k}
=\sum_{r=1}^R =  u_{i,r} v_{j,r} w_{k,r}$ makes this
optimization problem difficult.  
Alternatively,  
minimization of 
$|| \m Y - \v \Theta ||^2 +  \sum_{i,j,k,r}\lambda_r |u_{i,r} v_{j,r} w_{k,r} |$ or 
$|| \m Y - \v \Theta ||^2 +  \lambda_1 \sum_{i,r}|u_{i,r}| + 
  \lambda_2 \sum_{j,r}|v_{j,r}| + \lambda_3 \sum_{k,r}|w_{k,r}| $
would be feasible via modifications to the ALS procedure. 
The latter criterion 
provides estimators that are equivalent to posterior modes
under 
%multivariate versions of 
double exponential prior distributions, and 
%A penalization similar to the latter one has been used 
is similar to a criterion used
for matrix estimation by 
\citet{witten_tibshirani_hastie_2009}, as 
an alternative to least-squares estimation via the SVD. 
However, unlike a hierarchical modeling approach,
such $L_1$-penalized estimators always shrink towards zero, and do 
not take advantage 
of the potential variances and correlations in $\m Y$ 
(such as those described by Equation \ref{eq:hvar}) that could 
improve  estimation of $\v \Theta$.

Replication code and data for the numerical results in this paper are 
available at the author's website: 
\href{http://www.stat.washington.edu/~hoff}{\nolinkurl{http://www.stat.washington.edu/~hoff}}

\bibliographystyle{plainnat}
\bibliography{main}

\end{document}